# VizGen: Data Exploration and Visualization from Natural Language via a Multi-Agent AI Architecture

Sandaru Fernando
Department of Computational
Mathematics
University of Moratuwa
Moratuwa, Sri Lanka
fernandowsr.22@uom.lk

Imasha Jayarathne
Department of Computational
Mathematics
University of Moratuwa
Moratuwa, Sri Lanka
jayarathnekain.22@uom.lk

Sithumini Abeysekara
Department of Computational
Mathematics
University of Moratuwa
Moratuwa, Sri Lanka
abeysekarawmss.22@uom.lk

Shanuja Sithamparanthan
Department of Computational
Mathematics
University of Moratuwa
Moratuwa, Sri Lanka
shanujas.22@uom.lk

Thushari Silva
Department of Computational
Mathematics
University of Moratuwa
Moratuwa, Sri Lanka
thusharip@uom.lk

Deshan Jayawardana
Mitra AI
Moratuwa,Sri Lanka
djayawardana@mitrai.com

*Abstract—* Data visualization is essential for interpreting complex datasets, yet traditional tools often require technical expertise, limiting accessibility. VizGen is an AI-assisted graph generation system that empowers users to create meaningful visualizations using natural language. Leveraging advanced NLP and LLMs like Claude 3.7 Sonnet and Gemini 2.0 Flash, it translates user queries into SQL and recommends suitable graph types. Built on a multi-agent architecture, VizGen handles SQL generation, graph creation, customization, and insight extraction. Beyond visualization, it analyzes data for patterns, anomalies, and correlations, and enhances user understanding by providing explanations enriched with contextual information gathered from the internet. The system supports real-time interaction with SQL databases and allows conversational graph refinement, making data analysis intuitive and accessible. VizGen democratizes data visualization by bridging the gap between technical complexity and user-friendly design.

## I. INTRODUCTION

Data visualization plays a pivotal role in transforming raw datasets into actionable insights. It enables stakeholders across various sectors to identify hidden patterns, trends, and relationships that are not readily apparent in tabular or numerical formats. With the rapid acceleration of digital transformation, the volume, velocity, and variety of data have grown significantly, thereby increasing the demand for accessible and intelligent visualization tools.

Despite the proliferation of feature-rich platforms such as Microsoft Excel, Tableau, and Power BI, significant usability barriers persist. These platforms often require users to possess domain expertise, technical knowledge, and familiarity with visualization principles. The process of selecting suitable graph types, understanding variable mappings, and interpreting underlying data structures can be overwhelming, particularly for non-technical users. Consequently, the gap between data availability and meaningful insight extraction remains a critical challenge.

Recent advancements in Artificial Intelligence (AI) and Natural Language Processing (NLP) offer new opportunities to address these limitations. By leveraging AI-powered conversational interfaces, systems can now interpret user intent, analyze data semantics, and automate the creation of contextually appropriate visualizations with minimal user input.

In this work, we present VizGen, an AI-assisted system for natural language-driven graph generation. VizGen aims to democratize access to data visualization by allowing users to express their requirements in plain language, eliminating the need for technical or programming expertise. The objective of this study is to design and evaluate a system that bridges the gap between data complexity and user accessibility through intelligent, conversational graph generation and contextual insight explanation.

Our work makes several significant contributions to the field of automated data visualization. We propose a novel multi-agent framework consisting of specialized agents for SQL query generation, visualization recommendation, graph customization, insight extraction, and explanation generation. VizGen features a conversational interface capable of parsing user queries and translating them into structured database operations and visualization parameters. The system supports dynamic connectivity with multiple SQL-based databases, enabling real-time data retrieval and visualization updates. We introduce an intelligent recommendation mechanism that utilizes Large Language Models (LLMs) to suggest optimal chart types based on data properties and user intent. VizGen performs automated insight analysis to identify trends, anomalies, and correlations in the dataset, generating intuitive explanations by integrating external contextual data to enable users to make informed and actionable business decisions. Additionally, users can iteratively refine the generated visualizations using natural language commands, allowing intuitive, non-technical interaction and iterative improvement.

## II. RELATED WORK

The data visualization domain has transitioned from static graph plotting in spreadsheet applications to the interactive dashboards available in modern Business Intelligence (BI) platforms. Practitioners of the field are accustomed to early generation tools, such as Microsoft Excel, that provided limited charting abilities but also required practitioners to handle much of the underlying data manually and provided limited interactivity. This is the genesis of BI products like Power BI, Tableau, and Google Data Studio that all embraced the concept of dynamic dashboards with drilldowns, filters, and data connections. However, these products, while powerful, assume some level of technical knowledge of the dashboarding tools and require their users to manually specify the type of chart, define its filters, manage datasets such as removing duplicate data, and often combine these with time-consuming, manual processes. These assumptions and expectations pose a potential barrier to the development of intelligent and user-friendly data visualization systems.

Over the last two decades, tools and products for data visualization have evolved considerably. Early solutions such as Microsoft Excel enabled users to manually create charts, while modern platforms offer features such as automation, interactivity, and, in some cases, AI integration. However, many existing tools lack the ability to autonomously manage data connectivity, aggregation, and interpretation, which are increasingly critical for effective analysis. Recent advances in Natural Language Processing and Large Language Models have opened new opportunities for automated data visualization generation. Wu et al. [1] conducted an empirical study evaluating LLMs' potential in generating visualizations from natural language descriptions, demonstrating that transforming structured tabular data into sequential text prompts for LLM consumption is effective, particularly when considering table schema information. Building on these ideas, Narechania et al. [2] introduced NL4DV, a Python toolkit that processes a tabular dataset and a natural language query to produce an analytic specification comprising identified data attributes, analytic tasks, and corresponding Vega-Lite chart specifications enabling programmatic visualization generation. Additionally, most platforms offer limited flexibility and often do not support natural language interaction, intelligent chart recommendations, or automated insight generation. These limitations pose challenges for non-technical users seeking rapid and meaningful data exploration.

The emergence of conversational analytics represents a significant shift in how users interact with data systems. Natural Language Interfaces to Data (NLID) [3] have gained substantial attention from both research and industry communities. These interfaces enable users to query databases and generate visualizations using everyday language rather than technical commands. Microsoft Research has developed comprehensive frameworks for conversational data analytics that combine AI and NLP technologies to support natural language querying of relational databases, significantly lowering the barrier for general users to conduct data analysis. Similarly, Google Cloud's Conversational Analytics API, powered by Gemini, demonstrates the commercial viability of natural language data interaction by integrating NL2Query capabilities with context retrieval tools for accurate and relevant responses.

The Text-to-SQL domain has seen remarkable advancement with the integration of Large Language Models. Recent surveys by Zhu et al. [4] and Huang et al. [5] comprehensively review LLM-enhanced text-to-SQL generation, classifying approaches into prompt engineering, fine-tuning, pre-trained models, and agent-based groups according to training strategies. These developments have enabled more sophisticated natural language to database query translation, supporting real-time data retrieval and analysis. Multi-agent frameworks have emerged as particularly promising approaches for complex data analysis tasks. Recent work by researchers has demonstrated the effectiveness of PlotGen [6], a multi-agent LLM-based system for scientific data visualization that orchestrates multiple specialized agents for query planning, code generation, and multimodal feedback. This approach aligns with broader trends in multi-agent systems research, where collaborative LLM frameworks show significant improvements in complex problem-solving scenarios.

Currently, the most notable tools in the data visualization sector include Power BI, Microsoft's business analytics service allowing users to visualize data via dashboards and reports with a user-friendly drag-and-drop interface that allows users to take advantage of real-time data, but users still have to manually select their chart types and apply filters. Venngage serves as a web-based design tool with a specific focus on infographics and charts using templates that users can customize, excelling at static storytelling via visuals but lacking real-time database connection and capabilities for generating insights using intelligence. GraphMaker functions as an online lightweight charting tool that can quickly package structured data visually, though it has limited chart types, limited customizations or AI capabilities, and does not adjust visualizations based on user activity or intent cues. Amazon QuickSight operates as a BI service running through the cloud that leverages cloud computing and machine learning through anomaly detection, forecasting, and auto-narratives, however, it still requires configuration through dropdown selections or visual controls, and lacks the capability for natural language conversation and multi-agent reasoning.

Advanced AI-powered visualization tools are increasingly integrating natural language processing capabilities. Recent developments show that NLP-powered interfaces enable conversational chatbots, voice commands, and search queries where users can ask questions in plain language instead of complex database queries to receive visualized responses. Automated Machine Learning (AutoML) capabilities are enabling non-technical users to process data and generate interactive dashboards automatically by connecting data sources, while generative AI techniques focus on creating entirely new visualizations and dashboards customized for specific audiences and use cases. The integration of these technologies represents a significant advancement toward democratizing data visualization and making advanced analytics accessible to users across all skill levels.

While these tools have improved accessibility and interactivity in the data visualization space, they continue to lack truly conversational interfaces, intelligent customization, and automated insight generation from analytics. VizGen bridges this gap with a conversational, AI-driven data visualization pipeline. VizGen employs a multi-agent architecture to understand natural language requests, suggest appropriate charts based on context, customize visual aspects of charts, and derive useful insights, accomplishing these tasks in real time with minimal effort from the user.

TABLE I. FEATURES AND LEARNING CURVE COMPARISON

Note: A = Microsoft Power BI, B = Venngage, C = GraphMaker, D = Amazon QuickSight, E = VizGen.

| Feature | A | B | C | D | E |
|---|---|---|---|---|---|
| Automated Chart Selection | Part. | Yes | Yes | Yes | Yes |
| LLM-Based Recommendation | No | No | No | No | Yes |
| External Database Integration | Yes | No | No | Yes | Yes |
| Chatbot-Based Interface | No | No | Yes | Yes | Yes |
| Graph Customization via NL | No | No | No | No | Yes |
| AI-Based Predictive Analysis | No | No | No | Yes | Yes |
| Integration with External APIs for Contextual Analysis | No | No | No | No | Yes |
| Automated Insight Generation | Part. | No | No | Yes | Yes |

TABLE II. PERFORMANCE AND ACCESSIBILITY

| Performance & Accessibility | A | B | C | D | E |
|---|---|---|---|---|---|
| Learning Curve | High | Low | Low | Medium | Low |
| Technical Expertise Required | High | Low | Low | Medium | Low |
| Deployement Complexity | High | Low | Low | Medium | Medium |

## III. SYSTEM DESIGN

The core of the VizGen platform is a multi-agent system, built using LangChain and LangGraph, which allows it to handle complex user requests by breaking them down into specialized tasks [7], [8] The system's operational flow is orchestrated by a state-driven graph that dynamically routes a user's query through a series of interconnected agents, each responsible for a specific task [9].

### A. The Intent Classifier

The workflow [10] begins with a user's natural language query, which is first processed by the Intent Classifier. This node analyzes the input to determine the core intent(s) and routes the request to the appropriate agent(s). The classifier is designed to recognize and classify six distinct intents, which directly correspond to the agents in the system.

The Visualization intent is detected when the user's query explicitly requests to generate charts, graphs, or visual representations of data, such as "Show me a bar chart of sales by month," and routes the query to the Visualization Agent. The Insight intent is identified when the user's query asks to discover trends, anomalies, or patterns in the data, exemplified by queries like "Find any unusual spikes in website traffic," which routes the request to the Analysis Agent [9]. The Explanation intent is triggered when the user wants an explanation about discovered insights or data patterns, as demonstrated by requests such as "Explain why sales dropped in Q2," directing the query to the Explanation Agent.

The Customization intent addresses situations when the user wants to modify an existing visualization, handling requests like "Change the color of this chart to blue" through a dedicated Customizer node. The System intent manages system-level tasks such as "Connect to a new database," which is processed by a dedicated System node. Finally, the Other intent serves as a fallback category for queries that don't fit the above classifications.

This dynamic routing mechanism, governed by the intent classifier, ensures that only the necessary tools and agents are engaged, making the process highly efficient. For queries with multiple intents, such as "Show me a bar chart of sales and explain the biggest trend," the workflow will sequentially activate the necessary agents, including the SQL Agent, Visualization Agent, Analysis Agent, and Explanation Agent, to fulfill the entire request comprehensively.

### B. Specialized Agents and Their Tools

Each of the four primary agents is composed of one or more specialized tools (nodes) that perform specific functions within the workflow [10].

The SQL Agent serves as the foundation for all data retrieval operations within the system. This agent incorporates a Metadata Retriever [11] that fetches database schema information, providing essential structural details about the connected databases. The SQL Generator component translates the user's query and metadata into an optimized SQL query, ensuring efficient data retrieval. To maintain data integrity and performance, the SQL Validator verifies the correctness and efficiency of the generated SQL query before execution. Finally, the SQL Executor runs the validated SQL query on the connected database to fetch the raw data required for visualization.

The Visualization Agent handles all aspects of data visualization processing. It employs a Data Preprocessor that cleans and prepares the retrieved dataset, ensuring data quality and consistency for accurate visual representation. The Graph Ranker component analyzes the preprocessed data to recommend and rank the most suitable graph types based on data characteristics and visualization best practices [11].

The Analysis Agent focuses on autonomously analyzing data for insights and patterns. This agent utilizes an Insight Generator as its core tool for performing autonomous analysis on the data to identify patterns and trends based on the user's prompt. This component enables the system to automatically discover meaningful relationships and anomalies within the dataset without requiring explicit user guidance [9].

The Explanation Agent provides context and explanations for data insights discovered during the analysis process. It operates through three specialized components: the Insight Explanation Query Generator formulates search plans to gather external context relevant to the discovered insights, the Search Execution Engine executes the search plan to find relevant information from external sources, and the Explanation Generator synthesizes the search results with the data insights to create comprehensive explanations. This multi-step process ensures that users receive contextually rich explanations that enhance their understanding of the data patterns and their implications [12].

*C. Other Functional Nodes*

Beyond the four main agents, the system includes other crucial nodes that handle specific intents or finalize the workflow. The Customizer node is dedicated to handling customization requests, allowing users to modify existing visualizations through natural language commands such as changing colors, adjusting chart types, or modifying visual elements.

The System node manages system-level operations, including tasks such as connecting to new databases, managing user authentication, and exporting data in various formats. The Response Generator serves as the final node in the workflow, compiling the output from all preceding agents and presenting it to the user in a cohesive and user-friendly format that integrates visualizations, insights, and explanations into a unified response.

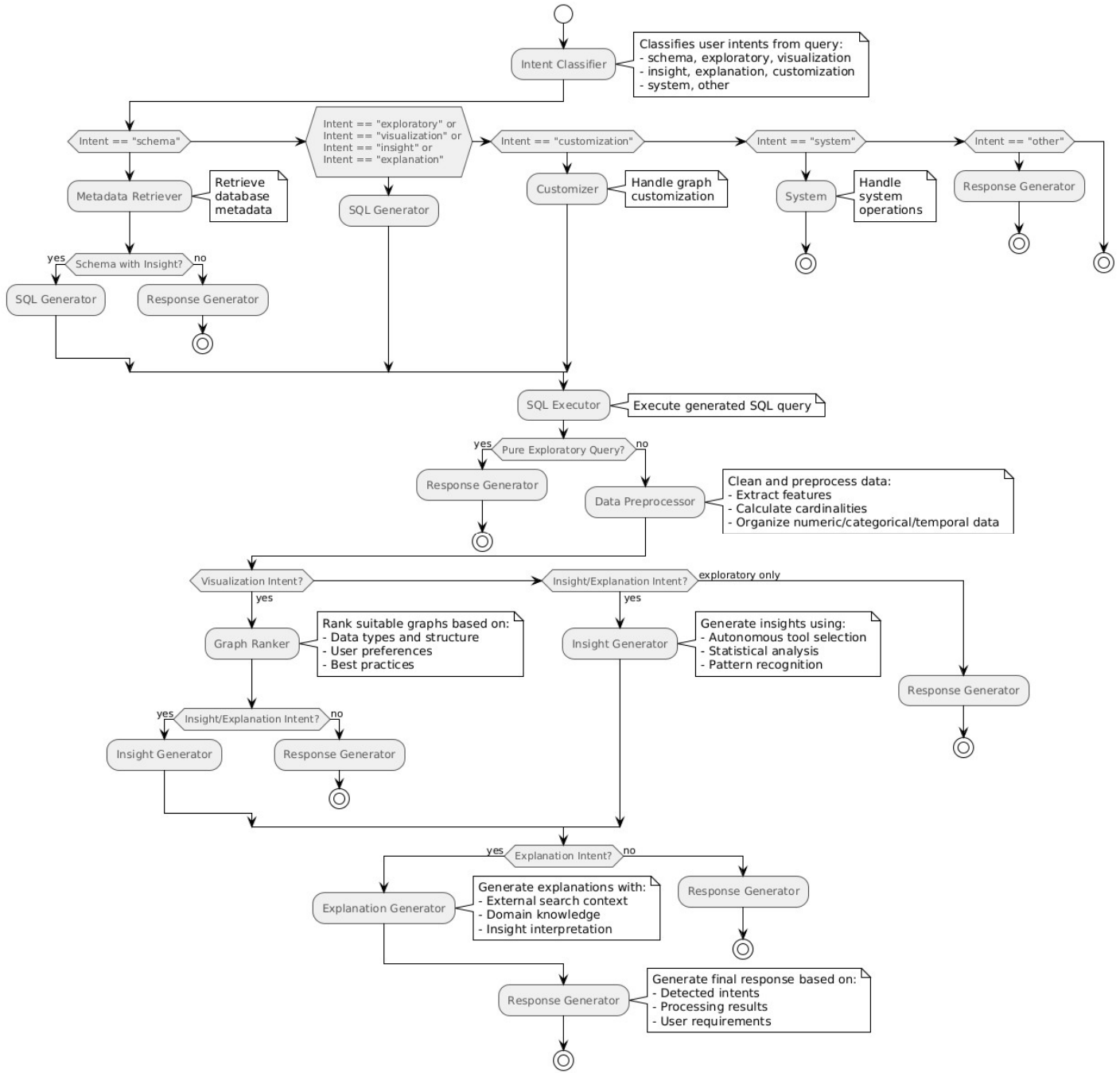

Fig. 1. Agentic Workflow of VizGen

## IV. Implementation

VizGen was implemented as a full-stack web application that integrated modern web frameworks, database systems and AI models.

Backend was developed using FastAPI because of its high performance, asynchronous capabilities, and seamless integration with Python-based AI and data libraries. Due to the system's dependence on LLM-driven dynamic query generation, FastAPI made it possible for rapid prototyping and effective API. MongoDB is selected as the primary database due to its flexibility in handling both unstructured and semi-structured data. As the system developed, it was advantageous that user interaction logs, visualization metadata, preferences, and feedback could all be stored without the need for rigid schemas. By integrating SQLAlchemy with external relational databases, the system supports user-provided connections to MySQL, PostgreSQL, MariaDB, MS SQL and Oracle DB.

On the frontend, React.js was used to build a responsive and modular interface. It was enhanced by Framer Motion to provide a polished user experience with seamless transitions and Material-UI for consistent design. The system interface included a natural language input area, dynamic graph visualization areas and controlling for filtering, zooming and exporting. Plotly.js allowed for the rendering various types of graphs, responsive to user input and backend responses.

One significant advancement was the incorporation of large language models (LLMs) for natural language understanding and SQL query generation, mainly Claude 3.7 Sonnet and Gemini 2.0 Flash. These models interpreted user queries, understood context, and generated suitable database queries. Furthermore, an LLM-based recommendation engine was introduced to suggest most suitable graph types based on data attributes and user intent.

The system was orchestrated using a multi-agent framework via LangChain, where each agent handling a distinct task, such as query generation, graph recommendation, and visualization rendering. During development, LangSmith was utilized for debugging and assessing LLM outputs, while LangGraph was utilized for branching workflows to handle edge cases and feedback loops.

GitHub was used for development and testing in a modular, structured codebase. The system was deployed on cloud platforms and services were containerized using Docker for testing scalability and accessibility. Automated testing tools (PyTest, Jest) and manual testing techniques guaranteed performance, usability, and dependability throughout implementation, paving the way for the system to move into evaluation and user testing phases.

## V. Evaluation

The VizGen tool was evaluated through a Google Forms survey comprising Likert-scale ratings (1–5) and open-ended questions. Nineteen people from a variety of backgrounds, including data analysts, developers, and undergraduate students, participated in the survey.

Participants' experience with data visualization tools ranged from beginner (level 1) to expert (level 5), with most users at intermediate to advanced levels:

TABLE III. EXPERIENCE LEVELS OF PARTICIPANTS

| Experience Level | Percentage |
|---|---|
| 1 (Beginner) | 5.3% |
| 2 | 10.5% |
| 3 | 31.6% |
| 4 | 36.8% |
| 5 (Expert) | 15.8% |

Commonly used tools among participants included Microsoft Excel (74%), Python visualization libraries (53%), and Power BI (21%)

### A. Key Ratings Summary

TABLE IV. SUMMRAY OF KEY RATINGS

| Evaluation Aspect | Average Rating | Key Insight |
|---|---|---|
| Clarity of Demo | 4.1 | Demo was clear and informative |
| Ease of Use | 4.4 | Users rated VizGen easy to use |
| Confidence Without Support | 4.4 | Most were comfortable using it by itself. |
| Overall UI Design | 4.4 | Positive comments about the user interface |
| Navigation Ease | 3.3 | Navigation needs to be improved |
| Speed and Responsiveness | 3.9 | Overall, good performance |

Average ratings were calculated for each evaluation aspect based on participant responses on a 1-to-5 Likert scale. The average rating was computed using the formula:

$$\text{Average Rating} = \frac{\sum_{i=1}^{N}(R_i)}{N}$$

where $R_i$ represents the rating given by the i[th] respondent, and N is the total number of respondents for that question.

### B. The most popular features

Data summary & insight (79%)
Auto chart recommendation (74%)
Text-to-Graph generation (68%)
AI chatbot support (68%)
Visual customization (53%)

All participants agreed that VizGen addresses a real need by simplifying complicated visualization tasks with natural language and AI automation. Projects involving data analysis (68%), business decision-making (16%), and research visualization (11%), were the main use cases found. Participants expect that VizGen will support a wide range of graph types, including time-series graphs, scatter plots, histograms, bar/column charts, line charts, and heatmaps.

To further improve usability and functionality of VizGen, participants proposed improvements such as database integration, CSV upload capability, support for 3D graphs, improved report generation, better chatbot voice interaction, and collaboration features.

In terms of adoption, 79% of participants indicated that they would be willing to participate in additional testing, and 95% of participants would suggest VizGen to others. Strong user interest and confidence in the tool were indicated by the average likelihood of future use, which was 4.4 out of 5.

In conclusion, the evaluation confirms VizGen as an effective, user-friendly, AI-driven graph generation tool that works well for users of all expertise levels. While the interface and feature set received high praise, navigation between features emerged as an area needing improvement. Overall, the outcomes show how promising VizGen is for streamlining data visualization processes in both professional and academic contexts.

## VI. Conclusion and Future Work

### A. Impact and Summary

VizGen represents a significant advancement in democratizing data visualization by successfully bridging the gap between complex data analytics and user accessibility. Through our multi-agent architecture and natural language interface, we have demonstrated that AI-powered systems can effectively automate the traditionally complex process of graph generation while maintaining high accuracy and user satisfaction. The system's ability to interpret natural language queries, generate appropriate SQL statements, and recommend suitable visualizations has proven effective across diverse user groups, from students to professional developers.

### B. Challengers Encountered

During development, several technical challenges emerged that required innovative solutions. External database connectivity proved particularly complex, with variations in driver compatibility and authentication methods necessitating tailored integration approaches for different database systems. The initial graph recommendation system produced inconsistent results, requiring extensive fine-tuning and incorporation of additional metadata analysis to achieve reliable performance.

Performance optimization presented another significant challenge, as the adoption of more accurate text-to-SQL models introduced latency issues that required careful balancing between accuracy and response time. Additionally, implementing natural language-based graph customization introduced edge-case handling complexities that demanded extensive debugging and model refinement.

### C. Future Research Directions

Several promising avenues for future development have emerged from our work. Expanding beyond SQL databases to include NoSQL systems such as MongoDB and document-based storage will significantly broaden the system's applicability. This expansion requires developing specialized query generation agents for different data paradigms while maintaining the intuitive natural language interface. Improving the accuracy and sophistication of graph type recommendations through enhanced machine learning models and user behavior analysis represents a key area for enhancement, including developing context-aware recommendation systems that consider user preferences, data characteristics, and visualization best practices simultaneously. Integrating support for real-time data streams and live dashboard updates would position VizGen as a comprehensive solution for dynamic data environments. This capability requires developing efficient data processing pipelines and optimized rendering mechanisms for continuous visualization updates. Developing native mobile applications and expanding platform support will make VizGen accessible across diverse computing environments, including optimizing the interface for touch interactions and ensuring consistent functionality across different screen sizes and operating systems. Furthermore, incorporating predictive analytics, anomaly detection, and statistical analysis capabilities directly into the visualization workflow would transform VizGen from a visualization tool into a comprehensive data analytics platform.

## Acknowledgment

The authors would like to express their gratitude to Prof. Thushari Silva of the Department of Computational Mathematics, University of Moratuwa, for her continuous guidance and valuable feedback throughout the development of VizGen. We also thank Mr. Deshan Jayawardana of Mitra AI for his industry insights and mentorship, which significantly contributed to the system's practical relevance. Our appreciation extends to the student and developer participants who took part in the evaluation study and provided constructive feedback. The support from the University of Moratuwa and Mitra AI in terms of resources and technical assistance is gratefully acknowledged.